\documentclass[
    ,final            
  ]
  {aipproc}

\layoutstyle{6x9}

\begin{document}

\title{Cosmological density perturbations in modified gravity theories}



\author{A. de la Cruz-Dombriz}{
  address={Departamento de F\'isica Te\'orica I, Universidad Complutense de Madrid, 28040 Madrid, SPAIN}}
\author{A. Dobado}{
  address={Departamento de F\'isica Te\'orica I, Universidad Complutense de Madrid, 28040 Madrid, SPAIN}}
\author{A. L. Maroto}{
  address={Departamento de F\'isica Te\'orica I, Universidad Complutense de Madrid, 28040 Madrid, SPAIN}
}
\begin{abstract}
In the context of $f(R)$ theories of gravity,
we study the cosmological evolution of scalar perturbations by 
using a completely general procedure. We find that the exact fourth-order 
differential equation for the matter density perturbations in 
the longitudinal gauge, reduces to a second-order
equation for sub-Hubble modes. This simplification is compared with the standard (quasi-static) equation used in the literature. We show that for general $f(R)$ functions the quasi-static approximation is not justified. However for those $f(R)$ adequately describing the present phase of accelerated expansion and satisfying local gravity tests, it does give a correct description for the evolution of perturbations.
\end{abstract}

\maketitle


\section{Introduction}
The present phase of accelerated expansion of the universe \cite{SN} poses one 
of the most important problems of modern cosmology. It is well known that
ordinary  Einstein's equations in either a matter or radiation dominated universe
give rise to decelerated periods of expansion. In order to have acceleration, 
the total energy-momentum tensor appearing on the right hand side of
the equations should be dominated at late times by a hypothetical negative 
pressure fluid  usually called dark energy \cite{review}.

However, there are other possibilities to generate a period of acceleration 
where Einstein's gravity itself is modified. 
In one of such possibilities,  new functions
of the scalar curvature ($f(R)$ terms)  are included  
in the gravitational action. Although such theories are able to describe
the accelerated expansion on cosmological scales correctly, they typically 
give rise to strong effects on smaller scales, but nevertheless, viable models can be constructed imposing strong constraints over $f(R)$ \cite{Silvestri_09_2007}:  $f_{RR}>0$ for high curvatures, $1+f_{R}>0$ for all $R$, $f_{R}<0$ to ensure that ordinary General Relativity behaviour is recovered at early times and $\vert f_{R}\vert\ll 1$  at recent epochs to ensure that local gravity tests hold. $f_{R}$ and $f_{RR}$ mean first and second derivative respectively of $f(R)$ with respect to the argument $R$.

The important question that arises is therefore  
how to discriminate dark energy models from modified gravities using observations.
It is known that by choosing particular $f(R)$ functions, 
one can mimic any expansion history, and in particular that of $\Lambda$CDM.
Accordingly, the exclusive use of observations from SNIa \cite{SN}, baryon acoustic
oscillations \cite{BAO} or CMB shift factor \cite{WMAP}, 
based on different distance measurements which are sensitive only to the expansion history, cannot settle the question of the nature of dark energy \cite{Linder2005}. 

However, there exists a different type of observations  which are sensitive, not only to the expansion history, but also to the evolution of matter density perturbations. The fact that the evolution of perturbations depends on the specific gravity model, i.e. it differs in general from that of Einstein's gravity even though the background evolution is the same, means that this kind of observations will help distinguishing between different models for acceleration.

In this work \cite{Maroto&Dombriz} we find the exact equation 
for the evolution of matter density perturbations for arbitrary $f(R)$
theories. Such problem had been previously considered in the literature \cite{perturbations} and approximated equations have been widely used based on the so called quasi-static approximation in which  all the  time derivative
terms for the gravitational potentials are discarded, and only those including density perturbations are kept. From our exact result, we will be able to determine under which conditions such an approximation can be justified. 
\section{Density perturbations evolution in $f(R)$ theories}
Let us consider the modified gravitational  action:
\begin{equation}
S=\frac{1}{16\pi G}\int d^4x \sqrt{-g}\left(R+f(R)\right)
\label{action}
\end{equation}
The corresponding equations of motion for that action are modified with respect to their usual General Relativity counterpartners. In our research we use a flat Robertson-Walker perturbed metric in the longitudinal gauge
\begin{equation}
ds^2\,=\,a^2(\eta)[(1+2\Phi)d\eta^2-(1-2\Psi)(dr^2+r^2d\Omega_{2}^2)] 
\label{perturbed_metric}
\end{equation}
where $\Phi\equiv\Phi(\eta,\overrightarrow{x})$ and 
$\Psi\equiv\Psi(\eta,\overrightarrow{x})$ are the scalar perturbations. Using this perturbed metric and a perturbed energy-momentum tensor for dust matter perfect fluid, the first order perturbed equations may be written for a general $f(R)$ function assuming that background equations hold. A relevant point of these theories is that potentials $\Phi$ and $\Psi$ are not equal provided $f_{RR}\neq0$.
From Einstein's equations, it is possible to derive a fourth order differential equation 
for matter density perturbation $\delta\equiv\delta\rho/\rho_{0}$ - with $\rho_{0}$ unperturbed density and $\delta\rho$ density constrast with respect to the background -alone. The process can be found in \cite{Maroto&Dombriz}. The resulting equation can be written as follows:
\begin{equation}
\beta_{4,f}\delta^{iv}+\beta_{3,f}\delta'''+(\alpha_{2}+\beta_{2,f})\delta''+(\alpha_{1}+\beta_{1,f})\delta'+(\alpha_{0}+\beta_{0,f})\delta \,=\,0
\label{delta_equation_separated}
\end{equation}
where the coefficients $\beta_{i,f}$  $(i\,=\,1,...,4)$ involve terms with $f_{R}'$ and $f_{R}''$. Equivalently,  $\alpha_{i}$ $(i\,=\,0,1,2)$ contain terms coming from the linear part (Einstein-Hilbert) of $f(R)$ in $R$ (background curvature). Parameter $\epsilon \equiv \mathcal{H}/k\equiv a'/(a k)$, where $'$ denotes derivative with respect to conformal time $\eta$, will be useful to perform a perturbative expansion of the previous coefficients $\alpha$'s and $\beta$'s in the sub-Hubble limit. We also define the following dimensionless parameters: $\kappa_{i} \equiv \mathcal{H}^{'^{(i)}}/\mathcal{H}^{i+1}$ 
and $f_{i}\equiv f_{R}^{'^{(j)}}/(\mathcal{H}^{j} f_{R})$ ($i, j=1,2,3$).
\section{Sub-Hubble modes and the quasi-static limit} 
We are interested in the possible effects on the growth of 
density perturbations once they enter the Hubble radius in the matter dominated era. 
In the sub-Hubble limit $\epsilon\ll 1$, it can be seen that the 
$\beta_{4,f}$ and $\beta_{3,f}$ coefficients are supressed by $\epsilon^2$
with respect to  $\beta_{2,f}$, $\beta_{1,f}$ and $\beta_{0,f}$, i.e., 
in this limit the equation for perturbations reduces to the 
following  second order expression:
\begin{equation}
\delta''+\mathcal{H}\delta'+\frac{(1+f_{R})^{5} \mathcal{H}^{2}
(-1+\kappa_1)(2\kappa_1-\kappa_2)-\frac{16}{a^8}f_{RR}^{4}(\kappa_2-2)k^{8}8\pi G \rho_{0}a^2}
{(1+f_{R})^{5}(-1+\kappa_{1})+\frac{24}{a^8}f_{RR}^{4}(1+f_{R})(\kappa_{2}-2)k^{8}}\delta \,=\,0
\label{eqn_ours}
\end{equation}
where we have taken only the leading terms in the $\epsilon$ expansion for
the $\alpha$ and $\beta$ coefficients. This expression can be compared with that usually  
considered in literature, obtained after performing strong 
simplifications in the perturbed equations by 
neglecting time derivatives of $\Phi$ and $\Psi$ potentials. 
Thus in \cite{Qstatic_Tsujikawa} they obtain:
\begin{equation}
\delta^{''}+\mathcal{H}\delta^{'}-\frac{1+4\frac{k^2}{a^2}
\frac{f_{RR}}{1+f_{R}}}{1+3\frac{k^2}{a^2}\frac{f_{RR}}{1+f_{R}}}
\frac{4\pi G\rho_{0}a^2 \delta}{1+f_{R}} \,=\,0
\label{Qstatic_strong_equation}
\end{equation}
We will now restrict ourselves to models satisfying all the viability conditions mentioned in the introduction, including $\vert f_{R}\vert\ll 1$.

Note that the quasi-static  expression \eqref{Qstatic_strong_equation} 
is only recovered in the matter era (i.e. for $\mathcal{H}=2/\eta$) or for
a pure  $\Lambda$CDM evolution for the background dynamics. 
Nevertheless in the limit $\mid f_{R}\mid\ll1$ 
it can be proven using the background equations of motion that
$1+\kappa_1-\kappa_2 \approx 0$ and therefore 
$2\kappa_1-\kappa_2 \approx -2+\kappa_2 \approx -1 
+ \kappa_1$ what is nothing but the fact that for viable models 
the background evolution resembles that of $\Lambda$CDM. This this fact allows to simplify  expression \eqref{delta_equation_separated} to approximately become \eqref{Qstatic_strong_equation}.

In other words, although for general $f(R)$ functions the quasi-static approximation is not justified, for those viable functions describing the present phase of accelerated expansion and satisfying local gravity tests, it gives a correct description for the evolution of perturbations.

In order to check our results we proposed two particular $f(R)$ models to obtain solutions both using \eqref{eqn_ours}  and \eqref{Qstatic_strong_equation}. As commented before, for viable models the background evolution resemble that of $\Lambda$CDM at low redshifts  and that of a matter dominated universe  at high redshifts. Nevertheless the $f(R)$ contribution 
gives the dominant contribution for small curvatures and therefore it may explain the cosmological acceleration.

The two studied models are of the form $f(R) \,=\, c_{1} R^p$ (where units $H_{0}^2$ will be considered). According to the results presented in \cite{Amendola}, models of this type include both matter dominated and  late-time accelerated universe provided the parameters  satisfy $c_{1} < 0$ and $0 < p < 1$. 

The first model (left figure) will have parameters $c_{1}=-4$ and $p=0.63$ and it is not accomplishing condition $\mid f_{R}\mid\ll1$. Second model (right figure) will have $c_1 \,=\,-4.3$ and $p\,=\,0.01$ and it does accomplish condition $\mid f_{R}\mid\ll1$. For each model, we have compared our result with the standard $\Lambda$CDM and the quasi-static approximation \eqref{Qstatic_strong_equation}. We saw that for first model quasi-static and correct approximation do not give the same evolution but for the second one the quasi-static approximation gives a correct description for the evolution.
\\
\\
\begin{minipage}{0.5\textwidth}
\resizebox{6.0cm}{4.0cm} 
{\includegraphics{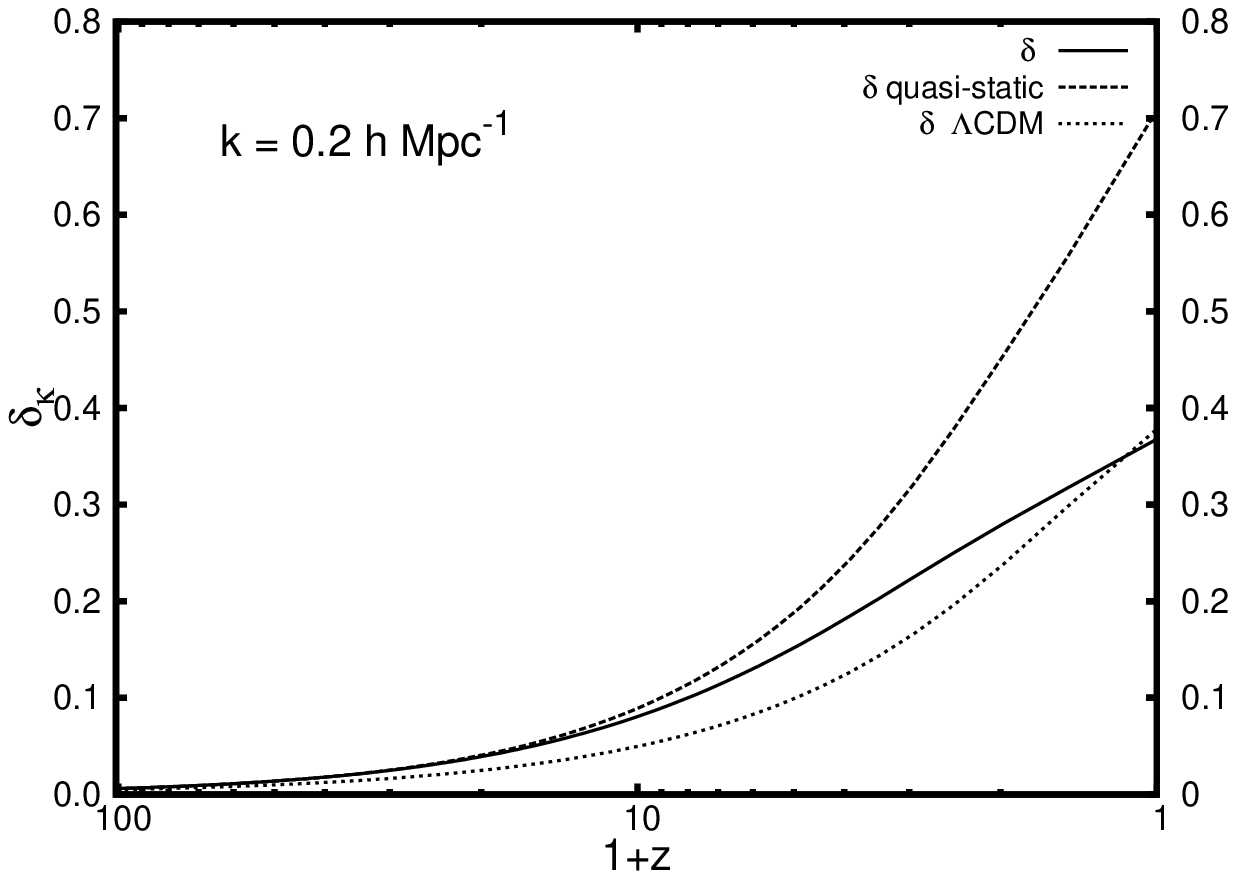}}\\
\end{minipage}%
\begin{minipage}{0.5\textwidth}
\resizebox{6.0cm}{4.0cm} 
{\includegraphics{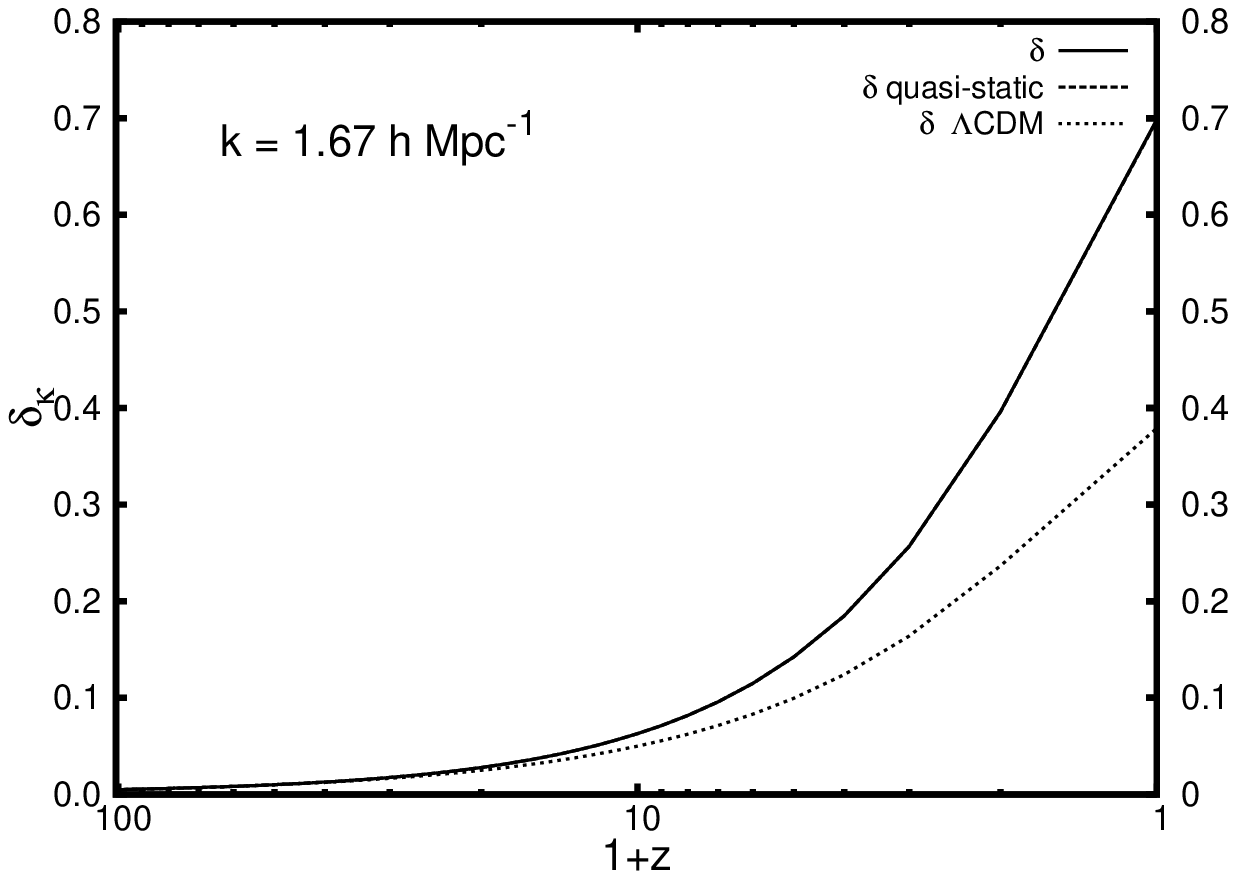}}\\
\end{minipage}
{\footnotesize $\bf{Left\,\,figure}$: $\delta_{k}$ with $k = 0.2 h\,Mpc^{-1}$ for first model and $\Lambda$CDM. Both, standard quasi-static evolution and equation \eqref{eqn_ours} have been plotted and they clearly differ.\,\,;\,\, $\bf{Right\,\,figure}$: $\delta_{k}$ with $k = 1.67 h\,Mpc^{-1}$ for second model and $\Lambda$CDM. The quasi-static evolution is indistinguishable from that coming from (\ref{eqn_ours}), but both diverge from $\Lambda$CDM behaviour as $z$ decreases.}
\section{Conclusions}
We have shown that for sub-Hubble modes, the differential equation for the evolution of density perturbation reduces to a second order equation and compared this result with that obtained within quasi-static approximation  used in the literature and found that for arbitrary $f(R)$ functions, such an approximation is not justified. 

However for theories with $\vert f_R\vert \ll 1$ today, perturbative calculation for sub-Hubble modes requires to take into account, not only the first terms, but also higher-order terms in $\epsilon={\cal H}/k$. In that case, the resummation of such terms modifies the equation which can be seen to be equivalent to the quasi-static case, but only if the universe expands as in a matter dominated phase or in a $\Lambda$CDM model. Finally, the fact that for models with $\vert f_R\vert \ll 1$ the background behaves today
precisely  as that of  $\Lambda$CDM makes the quasi-static approximation
 correct in those cases.
\vspace{.1cm}

{\bf Acknowledgements:} We would like to thank J.\,A.\,R.\,Cembranos and J.\,Beltr\'an.
 This work has been supported by the DGICYT (Spain) under projects FPA 2004-02602 and 2005-02327, CAM/UCM 910309 and by UCM-Santander PR34/07-15875.
\bibliographystyle{aipproc}   

\end{document}